\documentclass[twocolumn, showpacs,aps, pra,amsmath,amssymb,floatfix]{revtex4-1}
\usepackage{graphicx}
\usepackage{dcolumn}
\usepackage{bm}
\begin{document}
\title{Direct Photoassociative Formation of Ultracold KRb Molecules in the Lowest Vibrational Levels of the Ground State}
\author{Jayita Banerjee}
\email{jbanerjee@phys.uconn.edu}
\affiliation{Department of Physics, University of Connecticut, Storrs, Connecticut 06269-3046, USA}

\author{David Rahmlow}
\altaffiliation[ Presently at ]{Wyatt Technology Corporation, Santa Barbara, CA 93117-3253}
\affiliation{Department of Physics, University of Connecticut, Storrs, Connecticut 06269-3046, USA}

\author{Ryan Carollo}
\affiliation{Department of Physics, University of Connecticut, Storrs, Connecticut 06269-3046, USA}

\author{Michael Bellos}
\affiliation{Department of Physics, University of Connecticut, Storrs, Connecticut 06269-3046, USA}

\author{Edward E. Eyler}
\affiliation{Department of Physics, University of Connecticut, Storrs, Connecticut 06269-3046, USA}

\author{Phillip L. Gould}
\affiliation{Department of Physics, University of Connecticut, Storrs, Connecticut 06269-3046, USA}

\author{William C. Stwalley}
\affiliation{Department of Physics, University of Connecticut, Storrs, Connecticut 06269-3046, USA}
\date{\today}
\begin{abstract}
We report continuous direct photoassociative formation of ultracold KRb molecules in the lowest vibrational levels $(v''=0 -10)$ of the  electronic ground state $(X\,^1\Sigma^+)$, starting from $^{39}$K and $^{85}$Rb atoms in a magneto-optical trap. The process exploits a newfound resonant coupling between the $2(1),\,v'=165$ and $4(1),\,v'=61$ levels, which exhibit an almost equal admixture of the uncoupled eigenstates. The production rate of the $X^1\Sigma^+$ ($v''$=0) level is estimated to be $5\times10^3$ molecules/sec. 
\end{abstract}
%
%
%
\maketitle    
%
%
%
%
%
%
\section{Introduction}
Ultracold heteronuclear alkali dimers have enjoyed extensive attention from researchers in fields ranging from atomic, molecular and optical (AMO) physics to physical chemistry and chemical physics for over a decade now \cite{Book09, PCCP11}. The presence of a large molecule-fixed electric dipole moment in the absolute rovibrational ground state $(v''=0, J''=0)$ makes external control over the motion and internal quantum state very convenient, opening up the fields of ultracold chemical reactions and collisions \cite{Tscherbul06}, many body physics \cite{Uchler07} and quantum computation \cite{Rabi06}. To date, polar molecules in the $v =J =0$ level of the ground state have been successfully achieved by 1) magnetoassociation or photoassociation (PA) followed by stimulated Raman transfer in KRb \cite{Ospelkaus10,Aikawa10} and RbCs \cite{Sage05} and 2) photoassociation followed by direct radiative decay in LiCs \cite{Deiglmayr08} and NaCs \cite{Zabawa11}. Although the first technique can efficiently transfer a selectively prepared sample of vibrationally excited ultracold molecules to the lowest rovibronic level, the appeal of the second technique is that it provides a simple, single-step, continuous and irreversible process for converting ultracold atoms into molecules in the lowest rovibronic level.  

Ultracold KRb molecules, with a dipole moment of 0.66 Debye in the lowest rovibronic level, have been a favorite choice among the AMO community. In this letter we report a new single-step PA pathway for the formation of ultracold KRb molecules in the lowest vibrational levels of $X^1\Sigma^+$, $v''=0-10$, starting from laser cooled $^{39}$K and $^{85}$Rb atoms in a magneto-optical trap (MOT). PA paired with radiative decay has served as a very popular technique for producing ultracold molecules, normally in the highest vibrational levels of the electronic ground state \cite{Jones06}. However, to enhance the formation of ground-state molecules in vibrational levels far below the dissociation limit, it is possible to photoassociate to a rovibrational level of a single or a coupled excited electronic state with a radial probability distribution that includes both a long-range peak and a short-range peak. The long-range peak ensures efficient PA while the short range peak ensures efficient spontaneous radiative decay to low-lying levels. Such states occur if there is a strong resonant coupling between specific vibrational levels of two appropriate excited electronic states. This phenomenon has been observed in Rb$_2$ \cite{Pechkis07}, Cs$_2$ \cite{Dion01}, RbCs \cite{Gabbanini11} and is theoretically predicted for many heteronuclear dimers \cite{Banerjee10}. In NaCs \cite{Zabawa11}, this phenomenon is believed to lead the formation of molecules in the rovibronic ground state. In our present work, we have discovered a pair of states in KRb that are resonantly coupled, the 2(1) and 4(1) states. We perform extensive PA and resonance enhanced multiphoton ionization (REMPI) spectroscopy on the coupled states and the molecules formed from these states. These spectra have revealed the formation of ultracold KRb molecules in their lowest rovibrational levels. Our work presents a more complete experimental characterization of the resonant coupling of levels for formation of the lowest levels of the $X\,^1\Sigma^+$ ground state.  A complete theoretical understanding is of course also desirable, but difficult given the many coupled electronic states involved in KRb.

\begin{figure}
\centering \vskip 0 mm
\includegraphics[clip, width=\linewidth]{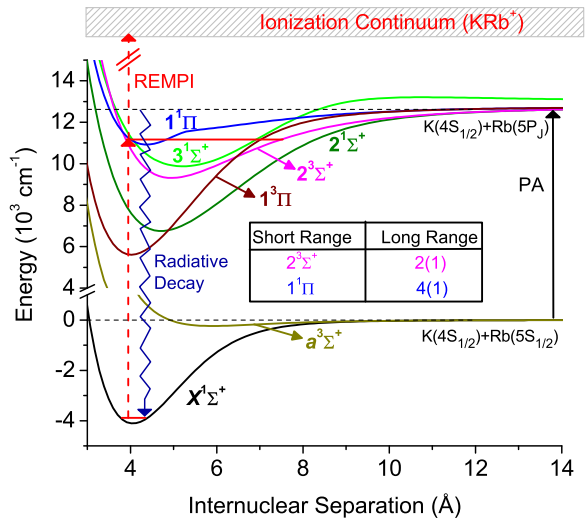}
\caption{\protect\label{fig1}(Color online)Schematic for formation and detection of ground-state KRb. The PA laser is detuned below the K$(4s)$ + Rb$(5p_{1/2})$ asymptote. The intermediate states used for detection of the lowest vibrational levels of the ground-state KRb molecules are $2\, ^3\Sigma^+$, $3\, ^1\Sigma^+$ and $1\, ^1\Pi$. At long-range, the $1\,^1\Pi$ state correlates to the 4(1) state and dissociates to the K$(4s)$ + Rb$(5p_{3/2})$ asymptote while the $2\, ^3\Sigma^+$ state diabatically correlates to the 2(1) and $2(0^-)$ states and dissociates to the K$(4s)$ + Rb$(5p_{1/2})$ asymptote.} 
\end{figure}

\section{Experiment}
A detailed description of the experimental apparatus is available in \cite{Wang04b}. Briefly, it consists of overlapped dual species MOTs of $^{39}$K and $^{85}$Rb, operating in a ``dark-SPOT'' configuration \cite{Ketterle93}, with atomic densities of $\sim3\times10^{10}$ cm$^{-3}$ and $\sim1\times10^{11}$ cm$^{-3}$ respectively. The temperatures of the K and Rb MOTs are on the order of $300\,{\mu}K$ and $100\,{\mu}K$ respectively. A cw Titanium:sapphire ring laser focused at the MOT with power of about 1 W acts as the PA laser. The molecules thus formed by PA then radiatively decay to the \emph{X} and \emph{a} states correlating with two ground-state atoms and are detected via REMPI. The REMPI laser is a Continuum model ND6000 pulsed dye laser and produces 10 ns pulses of $\sim$1 mJ energy that are focused to a diameter (FWHM) of $\sim$ 0.76 mm at the MOT. The ions are then detected by a channeltron and the KRb$^+$ ion signal is distinguished from K$^+$, Rb$^+$, K$_2^+$ and Rb$_2^+$ ions by time-of-flight mass spectroscopy. PA spectra are obtained by scanning the cw laser while keeping the pulsed REMPI laser fixed on a particular resonance. Similarly REMPI spectra are obtained by scanning the pulsed laser while fixing the PA laser on a known resonance. Figure 1 shows a schematic representation of our present work. The excited state potentials are \emph{ab initio} potentials obtained from \cite{Rousseau00} and the \emph{X} and \emph{a} state potentials correlating with two ground-state atoms are obtained from the experimental work reported in \cite{Pashov07}.

\begin{figure}
\centering \vskip 0 mm
\includegraphics[clip, width=\linewidth]{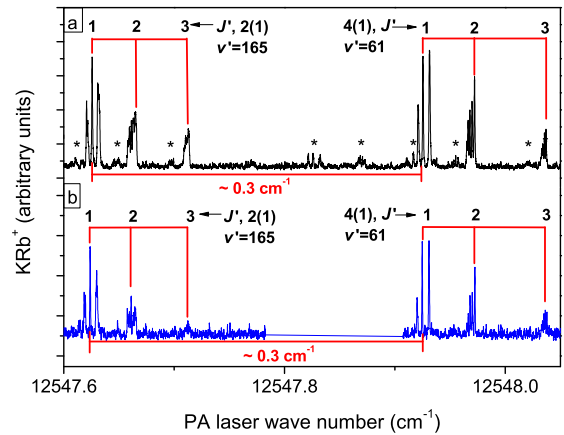}
\caption{\protect\label{fig3}(Color online)Photoassociation spectra of KRb with the detection laser tuned to transitions of a) $X\,^1\Sigma^+, (v''=84)$ to $4\,^1\Sigma^+, (v'=47)$ at 16635.8 cm$^{-1}$ and b) $X\,^1\Sigma^+, (v''=0)$ to $2\,^3\Sigma^+, (v'= 38)$ at 15116.14 cm$^{-1}$, showing the $2(1),\,v'=165$ and the $4(1),\,v'=61$ states. The lines marked by asterisks are $``$hyperfine ghost'' artifacts arising due to a small population in the bright hyperfine states (F=2 for $^{39}$K and F=3 for $^{85}$Rb) of $^{39}$K (4S) and $^{85}$Rb (5S) in our dark spot MOT. Hyperfine ghosts are present in spectrum b) also, but are not labeled to avoid congestion.} 
\end{figure}

\section{Resonant Coupling of the 2(1) and 4(1) states}
The coupled states used here can be observed by scanning the PA laser in the vicinity of 12547.9 cm$^{-1}$, red-detuned from the K$(4s)$ + Rb$(5p_{1/2})$ asymptote by 31 cm$^{-1}$. In the PA spectra of Figure 2, the two excited states are observed about $\sim$0.3 cm$^{-1}$ apart. These are identified as $v'=165$ of $2(1)$ and $v'=61$ of $4(1)$, using Hund's caes c) notation. In earlier work \cite{Wang04a}, these two states were observed using a pulsed REMPI laser with significantly broader linewidth and considerable amplified spontaneous emission, thus degrading the state selectivity but with the advantage of locating all PA resonances with some efficiency. In the present work PA spectra are obtained via REMPI using a narrow bandwidth (0.5 cm$^{-1}$) laser with a well-defined frequency. 

Theoretical calculations predict that the long-range 2(1), $v'=165$ level should radiatively decay predominantly to high-$v''$ levels of the \emph{X} state. Similarly the intermediate-range 4(1), $v'=61$ level should emit predominantly to low-$v''$ levels of the \emph{X} state. However, the PA spectra in Figure 2 reveal a different picture. Although Figure 2a corresponds to detection of high-$v''$ levels and Figure 2b to low-$v''$ levels of the ground-state, both have the same ratios of 2(1) and 4(1) signals for all the \emph{J} values. This suggests that the pair of excited states is strongly mixed by perturbative coupling.  

\begin{figure}
\centering \vskip 0 mm
\includegraphics[clip, width=\linewidth]{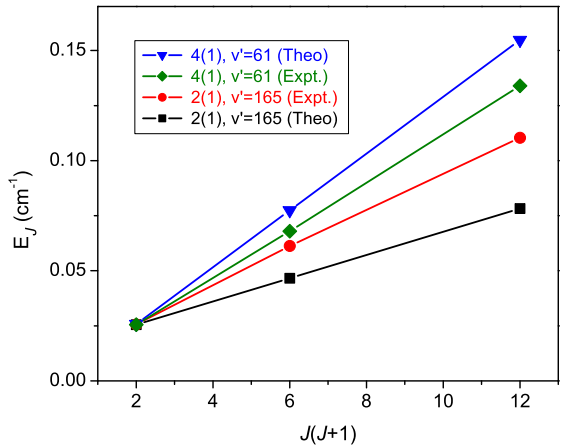}
\caption{\protect\label{fig3}(Color online) Plot of $E_J$ vs $J(J+1)$, where \emph{J} is the rotational quantum number and $E_{J}= J(J+1)B_{v}$. $B_v$ is the rotational constant. The theoretical values of $B_v$ are calculated using LEVEL \cite{LEV74} with single channel potentials calculated by Rousseau \emph{et al.} \cite{Rousseau00}.} 
\end{figure}

\begin{figure}
\centering \vskip 0 mm
\includegraphics[clip, width=0.9\linewidth]{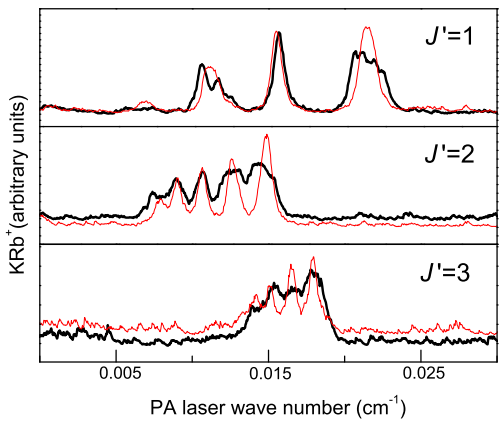}
\caption{\protect\label{fig4}(Color online)Comparison of hyperfine structure of the 2(1), $v'=165$ (black, thicker width) and 4(1), $v'=61$ (red, narrower width) states for $J'=1,\, 2$ and 3. The detection laser frequency for all the PA scans is 16635.8 cm$^{-1}$. Also, as the scan speeds and laser powers were not closely matched, we do not consider the differences in the linewidths to be significant.}
\end{figure}

Figure 3 depicts the behavior of the rotational constants for these two levels. The topmost (4(1), $v'=61$) and bottom-most(2(1), $v'=165$) lines depict theoretical prediction of $B_v$ based on single channel calculations. The experimental results depicted by the middle two lines deviate significantly towards one another, as expected in case of resonant coupling. The values of $B_v$ for 4(1), $v'=61$ and 2(1), $v'=165$ are quite similar and each deviates significantly from the single channel calculations or from simple linear interpolation between neighboring vibrational levels. This observation again suggests that these two excited states are strongly mixed. We also note that if they are strongly mixed, their hyperfine structure (hfs) should be quite similar. This is indeed observed, as can be seen in Figure 4 for superposed plots of the hfs for $J'=1,\, 2$ and 3.   

The vibrational numbering of the 4(1) electronic state is absolute. This state was previously observed by Kasahara \emph{et al.} \cite{Kasahara99} in an optical-optical double resonance experiment for higher values of \emph{J} and was later observed by our group \cite{Wang04b} for lower values of \emph{J}. However, the vibrational numbering of the long-range state of 2(1) is slightly uncertain. In previous work by our group \cite{Wang04b}, we observed many vibrational levels of 2(1) state, but not all. Based on a long-range extrapolation to the dissociation limit \cite{Wang04b}, we can definitively state that the 2(1) vibrational level  in the present experiment is the 22$^{nd}$ level below the dissociation limit. Calculating the vibrational levels of 2(1) state using the LEVEL program \cite{LEV74} and the single-channel potential \cite{Rousseau00}, we find that the energy of the $v'=165$ calculated level corresponds closely to our experimental vibrational level. Hence, for ease of discussion we will denote this vibrational level of the 2(1) state to be $v'=165$.

\begin{figure}
\centering \vskip 0 mm
\includegraphics[clip, width=\linewidth]{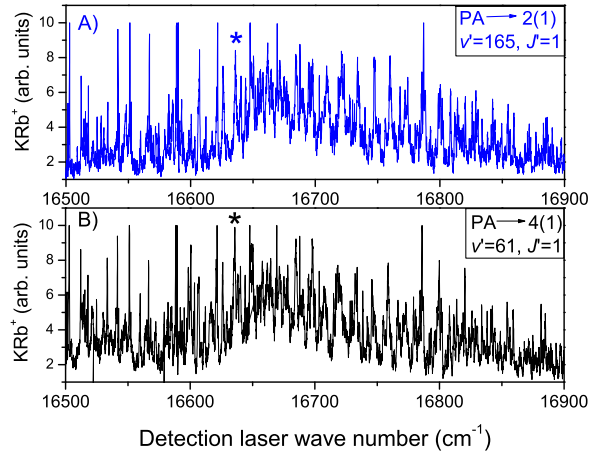}
\caption{\protect\label{fig5}(Color online)Near-identical REMPI spectra of KRb with the PA laser at A) 12547.6 cm$^{-1}$ and B) 12547.9 cm$^{-1}$. The line marked by an asterisk represents the detection frequency of 16635.8 cm$^{-1}$ used for the PA scans in Figure 2a and 4. Note that the intensity scales in A and B are the same.}
\end{figure}

\begin{figure*}
\centering \vskip 0 mm
\includegraphics[clip,width=\linewidth]{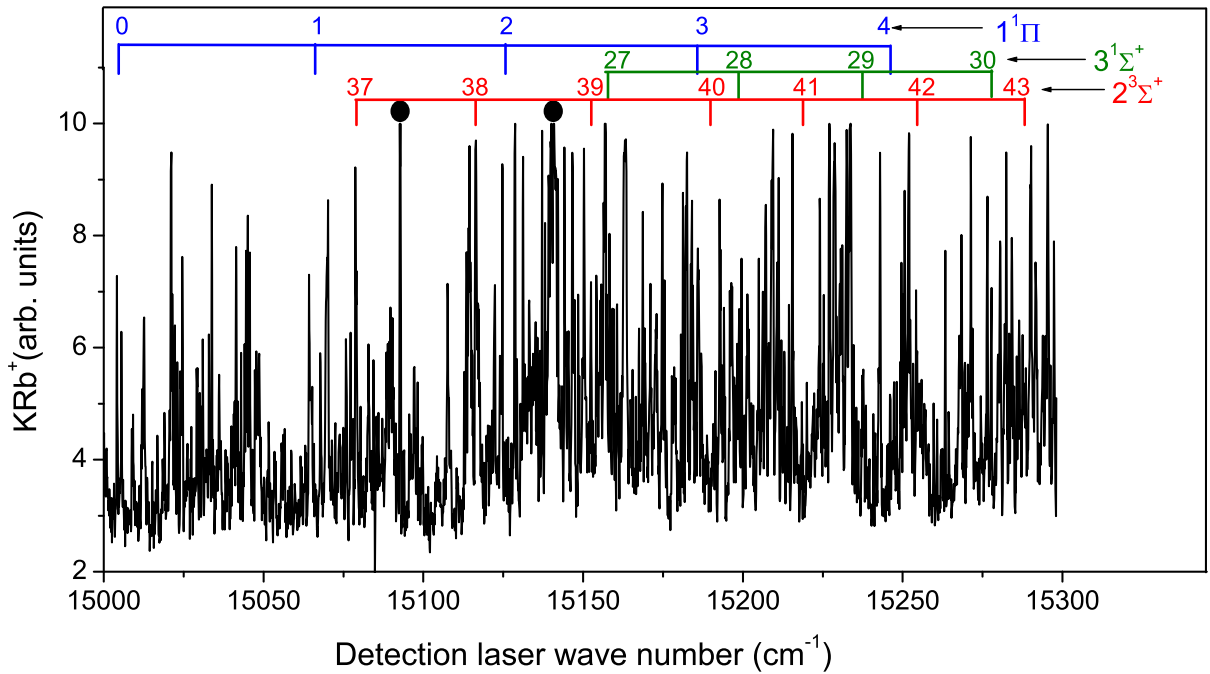}
\caption{\protect\label{fig6}(Color online)REMPI detection spectra of KRb molecules, at lower laser frequencies. The vertical lines indicate assignments from $X\,^1\Sigma^+, v''=0$ to several levels of the intermediate states $1\,^1\Pi, 2\,^3\Sigma^+$ and $3\,^1\Sigma^+$. The black circles ($\bullet$) indicate $^{85}$Rb two photon transitions that leak into the molecular channel. The PA laser was set to the 4(1), $v'=61, J'=1$ level for this scan.}
\end{figure*}

This mixing between 2(1) and 4(1) is even more apparent in the REMPI spectra of the \emph{X} and \emph{a} state molecules formed by radiative decay of the coupled PA states. These spectra are obtained by scanning the ionization laser with the PA laser fixed at 12547.62 cm$^{-1}\,(2(1), v'=165, J'=1$) and then at 12547.92 cm$^{-1}\,(4(1), v'=61, J'=1$), shown in Figure 5. These spectra are extremely congested and hence quite challenging to assign. Nevertheless comparing the two spectra reveals that they share the same spectral lines for the entire 400 cm$^{-1}$ scan except for six random peaks (out of $\sim$110 reproducible peaks). For many years, in our group, we have studied REMPI spectra, which are usually very different for different electronic states. We have never observed two REMPI spectra arising from two different electronic states with such striking similarity. This implies a strong coupling between the participating electronic excited states and suggests that these coupled states are nearly an equal admixture of the 2(1) and 4(1) electronic states at least in the range of internuclear separations for which radiative decay to bound levels of the ground state is significant. This degree of mixing has not been observed or predicted in prior work on ultracold molecules, as far as we know.

\section{Formation of the Lowest Vibrational Levels of the $X\,^1\Sigma^+$}
In an attempt to better understand the vibrational distribution of ground state molecules formed via these coupled excited states, we performed REMPI scans from $\sim$15000-15300 cm$^{-1}$, which is significantly below the scan range shown in Figure 5, since in that region the number of intermediate state resonances is fewer and hence the spectra are expected to be less complex.

\begin{figure}
\centering \vskip 0 mm
\includegraphics[clip, width=\linewidth]{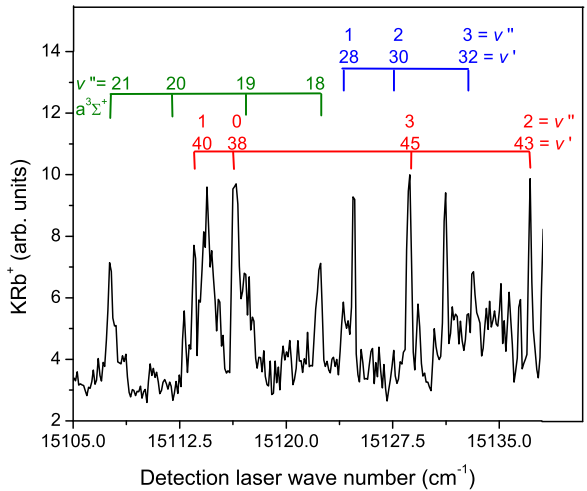}
\caption{\protect\label{fig7}(Color online)An expanded view of a selected region of the detection spectrum (Figure 6) of KRb molecules. The red vertical lines (bottom) indicate ground state molecules detected via the $2\,^3\Sigma^+$ state and the blue vertical lines (top) indicate those detected via the $3\,^1\Sigma^+$ state. The $v'$ corresponds to the intermediate state ($2\,^3\Sigma^+$ or $3\,^1\Sigma^+$ in the present case) vibrational levels and $v''$ corresponds to that of the ground $X\,^1\Sigma^+$ state. The green vertical lines (middle) correspond to the detection of molecules in $a\,^3\Sigma^+$ via the $3\,^3\Sigma^+$, $v'=11$ state. In the region of the spectrum shown in this figure nine lines (above the 4.5 vertical threshold) are assigned, while three are still unassigned.}
\end{figure}

The spectrum thus obtained, as shown in Figure 6, is still rather congested but nearly all of the lines have been successfully assigned. The assignments include numerous transitions from the lowest lying vibrational levels $v''=0-10$ of the $X\,^1\Sigma^+$ state, as well as from the levels $a\,^3\Sigma^+, v''=19-27$. In Figure 6, we have shown only the transitions from $X\,^1\Sigma^+, v''=0$ to the intermediate levels with the intention to avoid excessive congestion of assignments. The low-\emph{v} ground-state molecules, as observed by REMPI through the $1\,^1\Pi,\,2\,^3\Sigma^+$ and $3\,^1\Sigma^+$ excited electronic states, have been studied extensively in very recent molecular beam experiments \cite{JTKim11,JTKimPRA11, JTKim12}. Thus the vibrational energies of the ground state and intermediate states used to assign the REMPI spectra are all well known from experiments and hence there is very little uncertainty in the assignments of our spectra. Also, we have only assigned the lines in the REMPI spectrum which can be detected via the vibrational levels of the intermediate states already observed in \cite{JTKim11,JTKimPRA11, JTKim12}. As a result, there are a few lines that are still unassigned. 

In order to provide a cleaner view of the dense spectrum, we show in Figure 7 an expanded view of the spectrum in Figure 6 focusing on the region $\sim$15105-15137 cm$^{-1}$. In this region, lines can be assigned to transitions from $X\,^1\Sigma^+, v''=0-3$ levels to various vibrational levels of the $2\,^3\Sigma^+$ and $3\,^1\Sigma^+$ states. Transitions corresponding to the lowest triplet state are also shown. However, in this region, no distinct lines can be assigned to the $1\,^1\Pi$ state. As is evident from the figure, there are still some unassigned lines in this region. These are likely to be transitions to vibrational levels of the same intermediate states beyond the range reported in \cite{JTKim11,JTKimPRA11, JTKim12}. 

Further evidence that KRb molecules are formed in the lowest lying vibrational levels of the ground state comes from our observation of very large KRb$^+$ signals ionized using only a 532 nm pulsed laser. For molecules in high vibrational levels of the ground state, single-color ionization at 532 nm was previously observed to be quite inefficient. For $X\,^1\Sigma^+,\,v'' = 0$, the situation is much different because of a near resonance with the $3\,^3\Sigma^+,\,v' = 3$ level, which enhances two-photon ionization.  Because the vibrational spacings are very similar in the \emph{X} and the $3\,^3\Sigma^+$ states, other \emph{X}, low $v''$ levels are also enhanced. 

The relative intensities of the REMPI intermediate states are also of interest. The spin selection rule $\Delta S=0$ forbids transitions between singlet and triplet electronic states, so the 
$2\,^3\Sigma^+$$\leftarrow$$X\,^1\Sigma^+$ 
transitions might be expected to be very weak. However, as is evident in Figure 6, the 
$2\,^3\Sigma^+$$\leftarrow$$X\,^1\Sigma^+$ 
transitions are actually nearly as strong as the singlet-singlet REMPI transitions. Similarly efficient transitions between these two states are also reported in \cite{JTKim11,JTKimPRA11,JTKim12}. This occurs because there are five excited electronic states in this energy region, $2\,^1\Sigma^+, 1\,^1\Pi, 2\,^3\Sigma^+, 1\,^3\Pi$ and $3\,^1\Sigma^+$, which undergo mutual perturbations at various internuclear distances. This gives rise to very complicated coupling and hence the participating states are often not of pure singlet or triplet character. This is probably facilitating not only the detection of the lowest vibrational levels of $X\,^1\Sigma^+$ via the $2\,^3\Sigma^+$ state in KRb with very good efficiency \cite{JTKim11,JTKimPRA11}, but also the formation of these low-lying vibrational states by resonant coupling. At small internuclear distances the 2(1) state correlates to the $2\,^3\Sigma^+$ state and the 4(1) state correlates to $1\,^1\Pi$. Hence the long range states 2(1)$\sim$4(1) used for PA are themselves participants in this five-state perturbation complex, so that further study may reveal the influence of other nearby states on this strongly coupled pair. 

To estimate the formation rate of $X\,^1\Sigma^+,\,v''=0$ level via photoassociation, we calibrated the ion signal to determine the number of KRb$^+$ ions/pulse. Taking the velocity of the KRb molecules as $\sim$0.3 m/sec based on the MOT temperature of $\sim300\,\mu$K and estimating a detection volume of 0.45 mm$^3$ determined by the pulsed laser diameter of 0.76 mm and the nominal MOT thickness of 0.1 mm, the production rate of molecules in $v''=0$ state is approximately $5\times10^3$ molecules/second. Comparing with other recent results using single-step PA alone, this production rate is of the same order as the results \cite{Deiglmayr08} for LiCs and about a factor of ten smaller than for NaCs \cite{Zabawa11}.

The main caveat in using a single-step photoassociation process rather than a method of coherent transfer is the unwanted population of various higher vibrational levels due to spontaneous radiative decay. However a pure sample of rovibrational ground state molecules is still achievable after the PA process by several possible methods. As discussed in \cite{Viteau08}, vibrational cooling of the sample is one option. For KRb molecules, the $2\,^1\Sigma^+ - 1\,^3\Pi$ complex can provide intermediate states for optical pumping of molecules in higher vibrational levels down to $v''=0$. Another option for obtaining a pure sample of absolute ground state molecules  is to photodissociate the vibrationally excited molecules, so that the residue is a pure sample of $v''=0$. This process has been previously carried out successfully for K$_2$ \cite{Nikolov99}. Either method can be efficient, so the choice between them depends to some extent on the availability of suitable lasers.

\section{Summary}
In summary, we have presented a simple, single-step path for the efficient formation of KRb molecules in the $v''=0$ level of the $X\,^1\Sigma^+$ state via photoassociation through a pair of strongly resonantly coupled levels of the 2(1)$\sim$4(1) excited electronic states. The process forms $\sim5\times10^3$ KRb molecules/second in the $v''=0$ level of the ground state. It achieves this with just a single laser and works well at typical uncompressed MOT densities, with atomic temperatures of a few hundred $\mu$K. Further gains might be achieved by purifying the vibrational distribution into the $v''=0$ level and optically trapping them to allow studies of collisions and interactions. This process provides continuous production and accumulation of the absolute ground state molecules without complicated transfer techniques. Finally the observation of suitable pairs of resonantly coupled states for KRb, together with related observations in LiCs, NaCs and RbCs, inspires confidence that similar pathways can be found for other heteronulcear molecules as predicted in \cite{Banerjee10}. 

We gratefully acknowledge support from the National Science Foundation and the Air Force Office of Scientific Research (MURI).


%

\end{document}